\documentclass[conference,a4paper,final,twocolumn,10pt,twoside]{IEEEtran}
\usepackage{layout}
\IEEEoverridecommandlockouts 

\ifCLASSINFOpdf

\else
\fi
\usepackage{mathtools}
\usepackage{amsmath}
\usepackage{amssymb}
\usepackage{cite}
\usepackage{graphicx}
\usepackage{mathrsfs}
\usepackage{siunitx}
\usepackage{color}

\DeclareMathAlphabet{\mathscrbf}{OMS}{mdugm}{b}{n}
\usepackage{pdfpages}
\usepackage{graphicx}

\title{Two Quasi Orthogonal Space-Time Block Codes with
Better Performance and Low Complexity Decoder}
\author{\IEEEauthorblockN{Ali Lotfi-Rezaabad$^1 $, Siamak Talebi$^{2,3},$~\IEEEmembership{ Member,~IEEE}, and Ata Chizari$^1$}
\IEEEauthorblockA{$^1$Sharif University of Technology (SUT), Tehran, IRAN\\
$^2 $Advanced Communication Research Institute (ACRI), Sharif University of Technology (SUT), Tehran, IRAN \\$^3 $Shahid Bahonar University of Kerman (SBUK), Kerman, IRAN\\
Email: \{Lotfi-rezaabad\_Ali, Chizari.Ata\}@ee.sharif.edu, Siamak.Talebi@uk.ac.ir\\
}}

\begin{document}
\maketitle

\begin{abstract}
This paper presents two new space time block codes (STBCs) with quasi orthogonal structure for an open loop multi-input single-output (MISO) systems. These two codes have been designed to transmit from three or four antennas at the transmitter and be given to one antenna at the receiver. In this paper first, the proposed codes are introduced and their structures are investigated. This is followed by the demonstration of how the decoder decodes half of transmitted symbols independent of the other half. The last part of this paper discusses the simulation results, makes performance comparison against other popular approaches and concludes that the proposed solutions offer superiority.
\end{abstract}
\section{Introduction}
\IEEEPARstart{N}{owadays}, on the one hand explosion of data traffic in wireless communication is real and undeniable \cite{andrews2014will}, on the other hand, spectrum is the most valuable and scarce asset in communication. In order to achieve a high data transmission rate in narrow band wireless communication with an acceptable performance, we need to combat the interference between received symbols at the receiver which is called fading phenomenon in Rayleigh channels. One of the most advanced methods that is commonly deployed is space and time diversity which means the transmission of symbols not only from different places, but also at different times, requiring neither extra bandwidth nor more transmission power. The codes which benefit from these types of diversity are called space time block codes (STBCs).

Today, many STBCs with various characteristics designed for open loop multi-input multi-output (MIMO) systems are in existence. We can categorize some of codes into two groups, those with orthogonal and those with quasi orthogonal structures. Although orthogonal space-time block codes (OSTBCs) decoder is much simpler than quasi-orthogonal space-time block codes (QOSTBCs) one, the rate of QOSTBCs is higher than that of OSTBCs for more than two transmitter antennas. Let ponder on the STBCs which are presented in \cite{alamouti1998simple}, \cite{tarokh1999space} for quasi-static flat fading channels. An OSTBC with
very simple decoder and full rate is Alamouti code \cite{alamouti1998simple} and since of this feature it is a good choice for practical applications. Relaxing the simple decoding capability of OSTBCs, not being exist full rate OSTBCs for more than $2$
transmitter antennas is a drawback for these codes. To
overcome this shortcoming, QOSTBCs was introduced for the
first time in \cite{jafarkhani2001quasi} .
The presented code was for four transmitter antennas which can be approached to full rate at the expense of diversity.
Subsequently, both full rate and full diversity QOSTBC which was an
enhancement of \cite{jafarkhani2001quasi} appeared in \cite{tirkkonen2001optimizing}, \cite{sharma2002improved}. In both aforementioned studies, diversity has been improved from $2M$ to $4M$, in which $M$ is the total number of antenna on the receiver side, by constellation rotation for some symbols. The orthogonality space between the two groups of columns makes independent decoding of pairs of transmitted symbols possible.

Some codes with different structure from that of OSTBCs and QOSTBCs with dramatically better performance have also been reported. For instance, the code proposed in \cite{belfiore2005golden} known as Golden Code, is full rate, full diversity and also non-vanishing constant minimum determinate with $2\times2$ transmission matrices, likewise Perfect Codes with $3 \times 3, 4 \times 4,$ and $6 \times 6$ is studied in \cite{1683915} . The main advantage of these codes is that their rates are more than one, nonetheless this is at the expense of ML detection with high complexity and the requirement for more than two receiver antennas.

In this paper, we present two new STBCs, one for three and the other for four transmit antennas with quasi-orthogonal form. For more clarification, consider a codeword in which some groups of  its columns are orthogonal to the opposite groups' columns, while columns in the same group are not.  That was the reason that they have been called QOSTBCs. Due to the space created between column, we can decode half of transmitting symbols without dependence the other half.

These two codes are both full rate and full diversity. Note that full diversity can appear when some of the transmitted symbols are selected from a constellation and the others are selected from the rotation of that constellation. This derivation has been proven in both \cite{tirkkonen2001optimizing} and \cite{sharma2002improved}. In fact, the strategy which is used in this paper is to combat destructive fading coefficients. In other words, we have combined symbols through scarifying of neither full rate nor full diversity.

The rest of the paper is organized as follows; In section II
system model is introduced and in Section III structure of the
codes and the decoders are investigated. Simulation results are provided in Section IV. Finally, Section V concludes the paper.

\textit{Notation}: We used bold letter for matrices. Superscripts $(.)^H, ||.||_F,$ and $(.)^*$ to indicate Hermitian, Frobenius norm and
complex conjugation, respectively. In addition, we use 
$\mathbb{C}^{M\times N}$ to represent the set of $M\times N$ matrices over field of complex numbers.

\section{System Model}

\subsection{Transmission Model}
Let consider a quasi-static flat fading channel with $M_T$ and $M_R$ antennas at the transmitter and the receiver, respectively.
Throughout this paper, we assume that there is no information about the channel state on the transmitter side, but the receiver has informed about channel state information well. In addition they have been perfectly synchronized. Suppose that the transmitter sends the codeword $\textbf{C}\in\mathbb{C}^{T\times M_T} $ over $T$ time slot, thus we can model the system as below:

\begin{equation}\label{primaryequation}
\mathbf{Y} = \mathbf{CH}+\mathbf{N},
\end{equation}

in which $\mathbf{Y}\in\mathbb{C}^{T\times M_R}$ is the received signal, $\mathbf{H}\in\mathbb{C}^{M_T\times M_R}$ is fading
matrices and $\mathbf{N}\in\mathbb{C}^{T\times M_R}$ is the Additive White Gaussian Noise (AWGN). For a system that has been designed for two blocks
fading the following expression can be written as follows:

\begin{equation}\label{BlockFading}
\begin{bmatrix}
\mathbf{Y}_1 \\
\mathbf{Y}_2
\end{bmatrix}= 
\begin{bmatrix}
\mathbf{C}_1 & 0 \\ 0 & \mathbf{C}_2
\end{bmatrix}
\begin{bmatrix}
\mathbf{H}_1 \\ \mathbf{H}_2
\end{bmatrix}+
\begin{bmatrix}
\mathbf{N}_1 \\ \mathbf{N}_2
\end{bmatrix},
\end{equation}
where codeword can be defined as below:
\begin{equation}\label{CodeWord}
\mathbf{C}=\text{diag}\{\mathbf{C}_1,\mathbf{C}_2\}\in \mathbb{C}^{2T\times 2M_T}
\end{equation}

\subsection{Decoder Model}
For maximum-likelihood (ML) detection at the receiver, the
decoder examines all possible answers for this equation and
then decides on the minimum of the following equation:

\begin{equation}\label{decoder}
\hat{\mathbf{C}}=\arg\min_{\mathbf{C}^i}||\mathbf{Y}-\mathbf{C}^i\mathbf{H}||^2_F.
\end{equation}

\section{Proposed Method}
\subsection{Four Transmitter and One Receiver Antennas}
In this section, we present our two QOSTBCs and also
its proposed decoder. We begin with the QOSTBCs for $M_T=4$ and
$M_R=1$, and proceed with the analysis of ML decoder before
returning to consider the case of $M_T=3$ and $M_R=1$.
By using the Alamouti scheme \cite{alamouti1998simple} we can define $\pmb{\mathscr{G}}$ as matrix generator,
\begin{equation}\label{Alamouti}
\pmb{\mathscr{G}}(x_1,x_2)\triangleq
\begin{bmatrix}
x_1 & x_2 \\ -x_2^* & x_1^*
\end{bmatrix}.
\end{equation}
Note that it is obvious that the two columns of this matrix are orthogonal to each other. To construct $\mathbf{C}_1$ and $\mathbf{C}_2,$ we should write as follows:
\begin{equation}\label{C1}
\mathbf{C}_1\triangleq
\begin{bmatrix}
\pmb{\mathscr{G}}(S_1+jS_2) &  \pmb{\mathscr{G}}(\tilde{S}_3+j\tilde{S}_4) \\
-(\pmb{\mathscr{G}}(\tilde{S}_3+j\tilde{S}_4))^* & (\pmb{\mathscr{G}}(S_1+jS_2))^*
\end{bmatrix},
\end{equation}
and,
\begin{equation}\label{C2}
\mathbf{C}_2\triangleq
\begin{bmatrix}
\pmb{\mathscr{G}}(S_1-jS_2) &  \pmb{\mathscr{G}}(\tilde{S}_3-j\tilde{S}_4) \\
-(\pmb{\mathscr{G}}(\tilde{S}_3-j\tilde{S}_4))^* & (\pmb{\mathscr{G}}(S_1-jS_2))^*
\end{bmatrix},
\end{equation}
where $\tilde{S}_k=S_k\times e^{j\theta_k}$ for $k = 3,4,7,8$.
The aforementioned codewords are the outcome of a case where the careful combination of the two symbols results in the best possible performance. In this case, as Eq. (\ref{C1}) and Eq. (\ref{C2}), are showing we combine $S_k,$ and $S_{k+4}$, where
$k = 1, 2, \dots, 4$, orthogonally. After substituting (\ref{Alamouti}) in both (\ref{C1}) and (\ref{C2}), we can represent $\mathbf{C}_1$ and $\mathbf{C}_2$, as follows:
\begin{equation}\label{ComC1}
\mathbf{C}_1=
\begin{bmatrix}
S_1+jS_5 & S_2+jS_6 & \tilde{S}_3+j\tilde{S}_7&\tilde{S}_4+j\tilde{S}_8 \\
-S_2^*+jS_6^*  &  S_1^*-jS^*_5 & -\tilde{S}_4^*+j\tilde{S}_8^*& \tilde{S}_3^*-j\tilde{S}^*_7 \\
-\tilde{S}_3^*+j\tilde{S}_7^*  &  -\tilde{S}_4^*+j\tilde{S}^*_8 & S_1^*-jS_5^*& S_2^*-jS^*_6\\
\tilde{S}_4+j\tilde{S}_8  &  -\tilde{S}_3-j\tilde{S}_7 & -S_2-jS_6& S_1+jS_5
\end{bmatrix},
\end{equation}

\begin{equation}\label{ComC2}
\mathbf{C}_2=
\begin{bmatrix}
S_1-jS_5 & S_2-jS_6 & \tilde{S}_3-j\tilde{S}_7&\tilde{S}_4-j\tilde{S}_8 \\
-S_2^*-jS_6^*  &  S_1^*+jS^*_5 & -\tilde{S}_4^*-j\tilde{S}_8^*& \tilde{S}_3^*+j\tilde{S}^*_7 \\
-\tilde{S}_3^*-j\tilde{S}_7^*  &  -\tilde{S}_4^*-j\tilde{S}^*_8 & S_1^*+jS_5^*& S_2^*+jS^*_6\\
-\tilde{S}_4-j\tilde{S}_8  &  -\tilde{S}_3+j\tilde{S}_7 & -S_2+jS_6& S_1-jS_5
\end{bmatrix}.
\end{equation}
The final codeword can be concluded using Eq. (\ref{CodeWord}).

When a system utilize this QOSTBC (the quasi-orthogonality will be provided) structure, it  transmits $\mathbf{C}_1$ in the first four time slots from its four transmitter antennas, on the other side, the receiver receives $\mathbf{Y}_1,$ and buffer that, the second four timeslots is the time to send the next sub-codeword $\mathbf{C}_2,$ and on the receiver side an approximation of $\mathbf{C}_2$ ($\mathbf{Y}_2$), will be received. Note that the channel state while transmitting $\mathbf{C}_1,$ should be different from while the transmitter is transmitting $\mathbf{C}_2$.
This recent scenario can implemented by using a reconfigurable
antenna such as PIXEL antenna \cite{grau2007multifunctional} on the receiver side. PIXEL
antennas are capable to provide up to $5$ uncorrelated channel propagation states simultaneously, therefore even more diversity gain is achievable. It can be stated, therefore, that by employing the proposed strategy the channel changes from quasi static to block fading and as a result the probability of destructive fading effect, will be mitigated near to zero.

Given that eight symbols are transmitted in eight time slots, the code rate is one and because it employs one antenna on the receiver side (MISO structure), it is a full rate code. In order to achieve full diversity and to maximize the minimum of code gain distance (CGD), we have to select $S_k$, in which $k= 1, 2, \dots, 8$, from rotated constellation.
Let $\nu_k$ denotes the $k^{\text{th}}$ column of codeword, and then we can write:
\begin{equation}\label{Quasi}
\begin{aligned}
<\nu_1,\nu_i>=0,i\neq 1,4\quad &,\quad <\nu_5,\nu_i>=0,  i\neq 5,8, \\
<\nu_2,\nu_i>=0, i\neq 2,3\quad &,\quad <\nu_6,\nu_i>=0,  i\neq 6,7,
\end{aligned}
\end{equation}
in which $<\nu_i,\nu_j>=\sum_{\forall k} \nu_{ki} \nu^*_{kj} $, and $\nu_{ki}$ is refer to $k^{\text{th}}$ element from vector $\nu_i$.

The orthogonality between some columns made by $\pmb{\mathscr{G}}$ can help to simplify our ML decoder. As mentioned before, this type of coding creates block fading channels \cite{fazel2008space} which allows us to formulate an ML decoding equation as follows:

\begin{equation}\label{decoder2}
\begin{aligned}
\hat{\mathbf{C}}&=\arg\min_{\mathbf{C}^i}\sum_{L=1}^{2}||\mathbf{Y}_L-\mathbf{C}^i_L\mathbf{H}_L||^2_F\\
&=\arg\min_{\mathbf{C}^i}\sum_{L=1}^{2}\text{Tr}\{(\mathbf{C}^i_L \mathbf{H}_L)^H\mathbf{C}_L^i \mathbf{H}_L\}\\
&-2\mathcal{R}e[\text{Tr}\{(\mathbf{H}^H_L)^H(\mathbf{C}^i_L)^H\mathbf{Y}_L\}].
\end{aligned}
\end{equation}

On account of the quasi-orthogonal structure of the codeword, the above equation can be simplified into two independent parts. Due to space limitation, we proposed these two detection formulas on top of the next page.

In Eq. (\ref{decoder11}) and Eq. (\ref{decoder22}), $y_i\in\mathbf{Y}$ and similarly, $h_i\in\mathbf{H}, \forall i$. Since $F_1 (.) $ is independent from $F_2 (.) $, entirely, we can state that the receiver is capable to decode $ (S_1, S_4, S_5, S_8) $ and $ (S_2, S_3, S_6, S_7) $ separately. It means that the ML decoder is talented to minimize both Eq. (\ref{decoder11}) and Eq. (\ref{decoder22}) over all possible symbols. It is therefore clear that the complexity of the ML decoder for such proposed codeword and the system is $\mathcal{O}(m^4),$ instead of $\mathcal{O}(m^8)$. All the reasons for this dramatic mitigation is creative symbols combination in the proposed codeword. 

\begin{figure*}[!t]
\normalsize
\setcounter{equation}{11}
\begin{align}\label{decoder11}
\nonumber
F_1(S_1,\tilde{S}_4,S_5,\tilde{S}_8) =&(|S_1+jS_5|^2+|\tilde{S}_4+j\tilde{S}_8|^2)\sum_{i=1}^{4}|h_i|^2+(|S_1-jS_5|^2+|\tilde{S}_4-j\tilde{S}_8|^2)\sum_{i=5}^{8}|h_i|^2\\\nonumber + & 4\mathcal{R}e[(S_1+jS_5)(\tilde{S}_4+j\tilde{S}_8)]\times \mathcal{R}e[h_1 h_4^*-h_2 h_3^*]\\\nonumber
+ & 4\mathcal{R}e[(S_1-jS_5)(\tilde{S}_4-j\tilde{S}_8)]\times \mathcal{R}e[h_5 h_8^*-h_6 h_7^*]\\\nonumber -& 2\mathcal{R}e\bigg[\sum_{m=0}^{1}[S_{4m+1}(j)^m\sum_{k=0}^{1}(-1)^{mk}(y^*_{4k+1}h_{4k+1}+y_{4k+2}h^*_{4k+2}+y_{4k+3}h^*_{4k+3}+y^*_{4k+4}h_{4k+4})]\bigg]\\
+& 2\mathcal{R}e\bigg[\sum_{m=0}^{1}[\tilde{S}_{4m+4}(j)^m\sum_{k=0}^{1}(-1)^{mk}(y^*_{4k+1}h_{4k+4}-y_{4k+2}h^*_{4k+3}-y_{4k+3}h^*_{4k+2}+y^*_{4k+4}h_{4k+1})]\bigg]
\end{align}
\hrulefill 
\begin{align} \label{decoder22}
\nonumber
F_2(S_2,\tilde{S}_3,S_6,\tilde{S}_7) =&(|S_2+jS_6|^2+|\tilde{S}_3+j\tilde{S}_7|^2)\sum_{i=1}^{4}|h_i|^2+(|S_2-jS_6|^2+|\tilde{S}_3-j\tilde{S}_7|^2)\sum_{i=5}^{8}|h_i|^2\\\nonumber - & 4\mathcal{R}e[(S_2+jS_6)(\tilde{S}_3+j\tilde{S}_7)]\times \mathcal{R}e[h_1 h_4^*-h_2 h_3^*]\\\nonumber
- & 4\mathcal{R}e[(S_2-jS_6)(\tilde{S}_3-j\tilde{S}_7)]\times \mathcal{R}e[h_5 h_8^*-h_6 h_7^*]\\\nonumber -& 2\mathcal{R}e\bigg[\sum_{m=0}^{1}[S_{4m+2}(j)^m\sum_{k=0}^{1}(-1)^{mk}(y^*_{4k+1}h_{4k+2}-y_{4k+2}h^*_{4k+1}+y_{4k+3}h^*_{4k+4}-y^*_{4k+4}h_{4k+3})]\bigg]\\
+& 2\mathcal{R}e\bigg[\sum_{m=0}^{1}[\tilde{S}_{4m+3}(j)^m\sum_{k=0}^{1}(-1)^{mk}(y^*_{4k+1}h_{4k+3}+y_{4k+2}h^*_{4k+4}-y_{4k+3}h^*_{4k+1}-y^*_{4k+4}h_{4k+2})]\bigg]
\end{align}
\hrulefill

\end{figure*}
\subsection{Three Transmitter and One Receiver Antennas}
By omitting the fourth column from both $\mathbf{C}_1$ and $\mathbf{C}_2$ in (\ref{ComC1}) and \ref{ComC2}, respectively, a full rate QOSTBC can be recommended for $M_T=3$ and $M_R=1$ as it is illustrated in (\ref{ComC13}) and (\ref{ComC23}),

\begin{equation}\label{ComC13}
\mathbf{C}'_1=
\begin{bmatrix}
S_1+jS_5 & S_2+jS_6 & \tilde{S}_3+j\tilde{S}_7\\
-S_2^*+jS_6^*  &  S_1^*-jS^*_5 & -\tilde{S}_4^*+j\tilde{S}_8^*\\
-\tilde{S}_3^*+j\tilde{S}_7^*  &  -\tilde{S}_4^*+j\tilde{S}^*_8 & S_1^*-jS_5^*\\
\tilde{S}_4+j\tilde{S}_8  &  -\tilde{S}_3-j\tilde{S}_7 & -S_2-jS_6
\end{bmatrix},
\end{equation}

\begin{equation}\label{ComC23}
\mathbf{C}'_2=
\begin{bmatrix}
S_1-jS_5 & S_2-jS_6 & \tilde{S}_3-j\tilde{S}_7\\
-S_2^*-jS_6^*  &  S_1^*+jS^*_5 & -\tilde{S}_4^*-j\tilde{S}_8^*\\
-\tilde{S}_3^*-j\tilde{S}_7^*  &  -\tilde{S}_4^*-j\tilde{S}^*_8 & S_1^*+jS_5^*\\
-\tilde{S}_4-j\tilde{S}_8  &  -\tilde{S}_3+j\tilde{S}_7 & -S_2+jS_6
\end{bmatrix}.
\end{equation}
 The same approach applies when considering full diversity and maximizing the minimum of CGD. After constructing codeword $\mathbf{C}'$ based on Eq. (\ref{CodeWord}), it is easy to
understand the following expressions,
\begin{equation}\label{Orthogonality2}
\begin{aligned}
&<\nu_1,\nu_i>=0,i\neq 1\quad &&,\quad <\nu_5,\nu_i>=0,  i\neq 4 \\
&<\nu_2,\nu_i>=0, i\neq 2,3\quad &&,\quad <\nu_6,\nu_i>=0,  i\neq 5,6 
\end{aligned}
\end{equation}
Given (\ref{Orthogonality2}), we have a quasi-orthogonal structure; Thus again we can
break ML into two independent parts almost similar to Eq. (\ref{decoder11}), and
Eq. (\ref{decoder22}). Thus it is again possible to reduce complexity of the ML decoder
from $\mathcal{O}(m^8),$ to $\mathcal{O}(m^4)$. Note that in order to minimize decoder
complexity we can also use sphere-decoding algorithm based on
\cite{hassibi2005sphere} for both codewords $\mathbf{C},$ and $\mathbf{C}'$. In this case, two independent
sphere-decoders can be used, one for detecting
$(S_1, S_4, S_5, S_8),$ and the other to detect $(S_2, S_3, S_6, S_7)$.

\section{Simulation Results and Discussion}

\begin{figure}
\centering
\includegraphics[trim={4cm 8.8cm 4cm 8.5cm},width=8cm,clip]{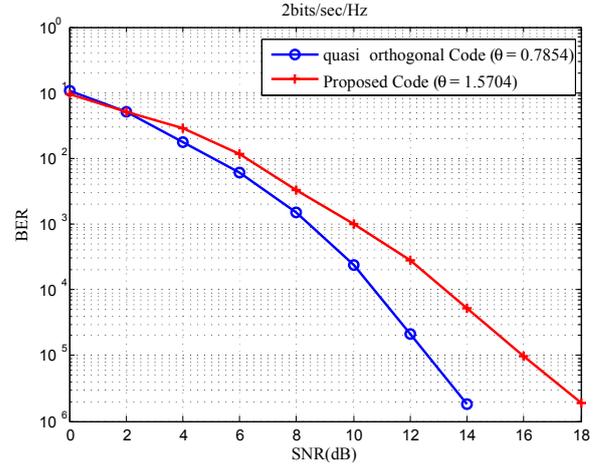}\\
\caption{Bit error rate versus signal to noise ratio for BPSK constellation
$1$ bit/sec/Hz ($ M_T = 4$).}
\label{4AntennaBPSK}
\end{figure}

\begin{figure}
\centering
\includegraphics[trim={4cm 8.8cm 4cm 8.5cm},width=8cm,clip]{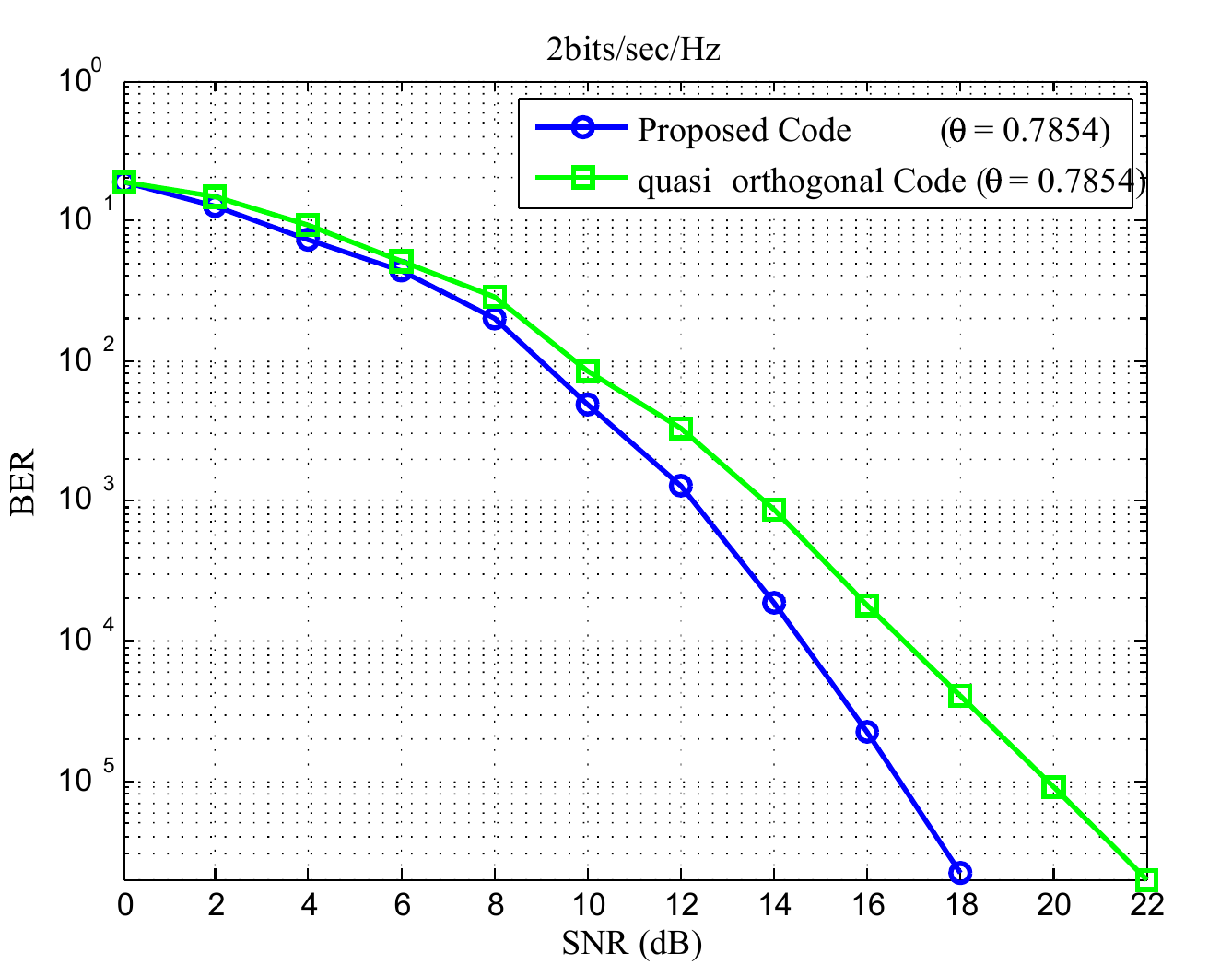}\\
\caption{Bit error rate versus signal to noise ratio for $4$QAM constellation
$2$ bits/sec/Hz ($ M_T = 4$).}
\label{4Antenna4QAM}
\end{figure}

\begin{figure}
\centering
\includegraphics[trim={4cm 8.8cm 4cm 8.5cm},width=8cm,clip]{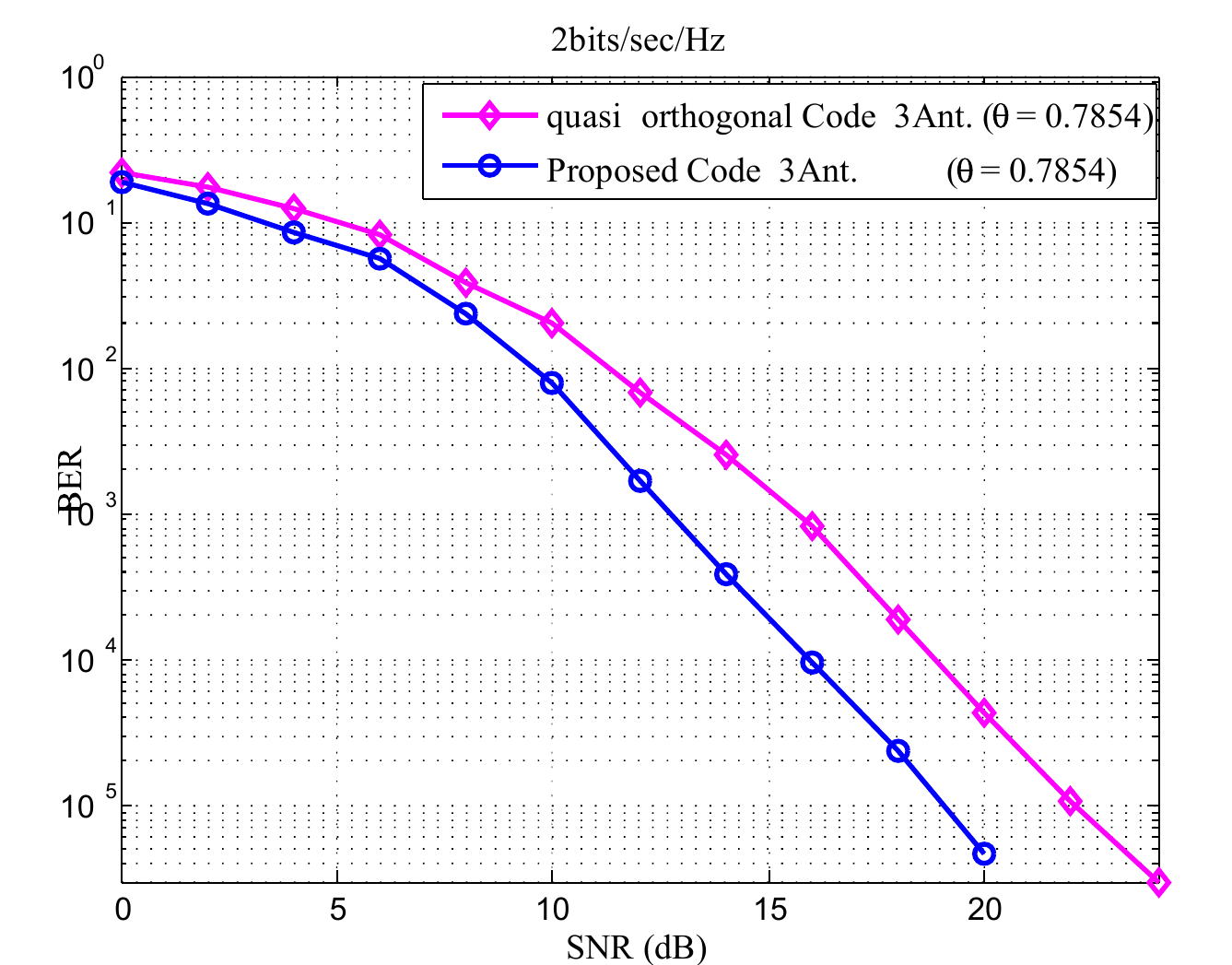}\\
\caption{Bit error rate versus signal to noise ratio for $4$QAM  constellation
$2$ bit/sec/Hz ($ M_T = 3$).}
\label{3AntennaQAM}
\end{figure}

In this part of the paper, we scrutinized the performance of our proposed codes and a popular approach namely QOSTBC.
Simulations were executed in an open loop MISO system using two separated sphere-decoding algorithm for detection and then the results were compared against quasi-orthogonal codes in \cite{tirkkonen2001optimizing}, \cite{sharma2002improved}, and \cite{jafarkhani2005space}. Fig. \ref{4AntennaBPSK} shows bit-error-rate (BER) versus signal to noise ratio (SNR) for the codeword $\mathbf{C}$, in which $M_T=4,$ and $M_R=1,$ using BPSK constellation. Likewise, Fig. \ref{4Antenna4QAM} depicts the results based on a $4$QAM constellation and
finally Fig. \ref{3AntennaQAM} represents the result for codeword $\mathbf{C}',$ where $M_T = 3,$
and $M_R=1,$ using similar constellation to Fig. 2. As simulations
for CGD shows, the optimal rotation is $\theta=\frac{\pi}{4},$ using aforementioned
constellation. With this amount of rotation we can reach the best performance for both two codewords. Consider that this code scheme is flexible and can be implemented with higher
order of QAM without any limitation.

By recent FPGA deployment, which provides higher speed for processing,  if we ignore complexity of the
decoder, the simulations were run under fair conditions. Therefore, according to the simulations, the proposed codes outperform those reported in \cite{sharma2002improved}
and \cite{jafarkhani2005space} by an amount near to $4$ dB.


\section{Conclusion and Future Work}
In this paper, we unveiled two full rates and full diversity quasi-orthogonal space-time block codes for open loop MISO systems employing three and four transmitter antennas and one antenna at the receiver. We demonstrated that ML decoder can be separated into two independent functions to reduce complexity of the decoder. By exploiting these codes, we changed property of the channel from quasi-static fading to block fading and discovered that this alteration improved the proposed codes' performance by up to $4$dB in comparison with \cite{sharma2002improved} and \cite{jafarkhani2005space}.

We can now pose this question: is it possible to reduce
complexity of the decoder for these codewords? Finding an
answer to this question is an open field problem for future work.
\bibliographystyle{IEEEtran}

\bibliography{IEEEabrv}
\end{document}